\documentstyle[12pt]{article}                    
\newfont{\ffont}{msym10}                        
\newcommand{\beq}{\begin{equation}}             
\newcommand{\eeq}{\end{equation}}               
\newcommand{\bqry}{\begin{eqnarray}}            
\newcommand{\eqry}{\end{eqnarray}}              
\newcommand{\bqryn}{\begin{eqnarray*}}          
\newcommand{\eqryn}{\end{eqnarray*}}            
\newcommand{\preprint}[1]{\begin{table}[t]      
            \begin{flushright}                  
            \begin{large}{#1}\end{large}        
            \end{flushright}                    
            \end{table}}                        
\newcommand{\PD}[2]                             
    {\frac{\partial^{#2}}{\partial #1^{#2}}}    
\begin{document}
\preprint{LA-UR-97-1494 \\ TAUP-XXXX-97}
\title{New Mass Relations \\ for Heavy Quarkonia}
\author{\\ L. Burakovsky\thanks{E-mail: BURAKOV@PION.LANL.GOV}, \
T. Goldman\thanks{E-mail: GOLDMAN@T5.LANL.GOV} \
\\  \\  Theoretical Division, MS B285 \\  Los Alamos National  
Laboratory \\
Los Alamos, NM 87545, USA \\  \\  and  \\  \\
L.P. Horwitz\thanks{E-mail: HORWITZ@TAUNIVM.TAU.AC.IL. Also at  
Department of
Physics, Bar-Ilan University, Ramat-Gan, Israel  } \
\\  \\ School of Physics and Astronomy \\ Tel-Aviv University \\  
Ramat-Aviv,
69978 Israel \\}
\date{ }
\maketitle
\begin{abstract}
By assuming the existence of (quasi)-linear Regge trajectories for heavy 
quarkonia in the low energy region, we derive a new, sixth power,  
meson mass
relation which shows good agreement with experiment for both  
charmed and beauty
mesons. This relation may be reduced to a quadratic  
Gell-Mann--Okubo type
formula by fitting the values of the Regge slopes of these  
(quasi)-linear
trajectories. For charmed mesons, such a formula holds with an  
accuracy of
$\sim $ 1\%, and is in qualitative agreement with the relation obtained 
previously by the application of the linear mass spectrum to a meson 
hexadecuplet.
\end{abstract}

\bigskip

\section*{  }
The generalization of the standard $SU(3)$ Gell-Mann--Okubo mass formula 
\cite{GMO} to higher symmetry groups, e.g., $SU(4)$ and $SU(5),$ became a
natural subject of investigation after the discovery of the fourth  
and fifth
quark flavors in the mid-70's \cite{disc}. Attempts have been made  
in the
literature to derive such a formula, either quadratic or linear in  
mass, by a)
using group theoretical methods \cite{Mac,BMO,Boal}, b) generalizing the 
perturbative
treatment of $U(3)\times U(3)$ chiral symmetry breaking and the  
corresponding
Gell-Mann-Oakes-Renner relation \cite{GOR} to $U(4)\times U(4)$  
\cite{GLR,MRT},
c) assuming the asymptotic realization of $SU(4)$ symmetry in the  
algebra
$[A_\alpha ,A_\beta ]=if_{\alpha \beta \gamma }V_{\gamma }$ (where  
$V_{\alpha
},A_{\beta }$ are vector and axial-vector charges, respectively)  
\cite{HO}, d)
extending the Weinberg spectral function sum rules \cite{Wei} to  
accommodate
the higher symmetry breaking effects \cite{BGN}, and e) applying  
alternative
methods, such as the linear mass spectrum for meson  
multiplets\footnote{Here,
we speak of linear spectrum over the multiplet quantum numbers,  
taking proper
account of degeneracy, not (directly) make use of linear Regge  
trajectories.}
\cite{lin,su4}. In the following\footnote{Here $\eta $ stands for  
the masses
of both isovector and isoscalar $n\bar{n}$ states which coincide on  
a naive
quark model level.}, $\eta ,\eta _s,\eta _c,\eta  
_b,K,D,D_s,B,B_s,B_c$ stand
for the masses of the $n\bar{n}$ $(n\equiv u$ or  
$d),s\bar{s},c\bar{c},b\bar{b
},s\bar{n},c\bar{n},c\bar{s},b\bar{n},b\bar{s},b\bar{c}$ mesons,
respectively\footnote{Since these designations apply to all spin states, 
vector mesons will be confusingly labelled as $\eta $'s. We ask the  
reader to
bear with us in this in the interest of minimizing notation.}. The  
linear mass
relations
\bqry
D & = & \frac{\eta +\eta _c}{2},\;\;\;D_s\;=\;\frac{\eta _s+\eta  
_c}{2}, \\
B & = & \frac{\eta +\eta _b}{2},\;\;\;B_s\;=\;\frac{\eta _s+\eta  
_b}{2},\;\;\;
B_c\;=\;\frac{\eta _c+\eta _b}{2}
\eqry
found in \cite{Boal,MRT}, although perhaps justified for vector mesons,
since a vector meson mass is given approximately by a sum of the  
corresponding
constituent quark masses, $$m(i\bar{j})\simeq m(i)+m(j)$$ (in fact,  
for vector
mesons, the relations (1),(2) hold with an accuracy of up to $\sim 4$\%),
are expected to fail for other meson multiplets, as confirmed by direct
comparison with experiment. Similarly, the quadratic mass relation
\beq
D_s^2-D^2=K^2-\eta ^2
\eeq
obtained in ref. \cite{GLR} by generalizing the $SU(3)$  
Gell-Mann-Oakes-Renner
relation \cite{GOR} to include the $D$ and $D_s$ mesons (here $\pi  
$ stands
for the mass of the $\pi ,$ etc.),
\beq
\frac{\pi ^2}{2n}=\frac{K^2}{n+s}=\frac{D^2}{n+c}=\frac{D_s^2}{s+c},
\eeq
(and therefore $D_s^2-D^2=K^2-\pi ^2\propto (s-n),$ also found in refs. 
\cite{BMO,BGN,HO}), does not agree with experiment. For  
pseudoscalar mesons,
for example, one has (in GeV$^2)$ 0.388 for the l.h.s. of (3) vs.  
0.226 for
the r.h.s.. For vector mesons, the corresponding quantities are   
0.424 vs.
0.199, with about 100\% discrepancy. The reason that the relation  
(3) does not
hold is apparently due to the impossibility of perturbative  
treatment of $U(4)
\times U(4)$ symmetry breaking, as a generalization of that of  
$U(3)\times U(3
),$ due to very large bare mass of the $c$-quark as compared to  
those of the
$u$-, $d$- and $s$-quarks. In ref. \cite{su4} by the application of  
the linear
spectrum to $SU(4)$ meson hexadecuplet, the following relation was  
obtained,
\beq
12\bar{D}^2=7\eta _0^2+5\eta_c^2,
\eeq
where $\bar{D}$ is the average mass of the $D$ and $D_s$ states  
(which are mass
degenerate when $SU(4)$ flavor symmetry is broken to $SU(3)),$ and
$\eta _0$ is the mass average of the corresponding $SU(3)$ nonet.  
As shown in
ref. \cite{su4}, this relation holds with an accuracy of up to  
$\sim $ 5\% for
all well established meson hexadecuplets.

It is well known that the hadrons composed of light $(u,d,s)$  
quarks populate
linear Regge trajectories; i.e., the square of the mass of a state with 
orbital momentum $\ell $ is proportional to $\ell :$ $M^2(\ell  
)=\ell /\alpha
^{'}+\;{\rm const,}$ where the slope $\alpha ^{'}$ depends weakly on the 
flavor content of the states lying on the corresponding trajectory,
\beq
\alpha ^{'}_{n\bar{n}}\simeq 0.88\;{\rm GeV}^{-2},\;\;\;
\alpha ^{'}_{s\bar{n}}\simeq 0.84\;{\rm GeV}^{-2},\;\;\;
\alpha ^{'}_{s\bar{s}}\simeq 0.80\;{\rm GeV}^{-2}.
\eeq
In contrast, the data on the properties of Regge trajectories of hadrons 
containing heavy quarks are almost nonexistent at the present time,  
although
it is established \cite{BB} that the slope of the trajectories  
decreases with
increasing quark mass (as seen in Eq. (6)) in the mass region of  
the lowest
excitations. This is due to an increasing (with mass) contribution  
of the
color Coulomb interaction, leading to a curvature of the trajectory  
near the
ground state. However, as the analyses show \cite{BB,KS,QR}, in the  
asymptotic
regime of the highest excitations, the trajectories of both light  
and heavy
quarkonia are linear and have the same slope $\alpha ^{'}\simeq  
0.9$ GeV$^{-2
},$ in agreement with natural expectations from the string model.

Knowledge of Regge trajectories in the scattering region, i.e., at  
$t<0,$ and
of the intercepts $a(0)$ and slopes $\alpha ^{'}$ is also useful  
for many
non-spectral purposes, for example, in the recombination \cite{rec} and 
fragmentation \cite{fra} models. Therefore, as pointed out in ref.  
\cite{BB},
the slopes and intercepts of the Regge trajectories are the fundamental 
constants of hadron dynamics, perhaps generally more important than  
the mass
of any particular state. Thus, not only the derivation of a mass  
relation but
also the determination of the parameters $a(0)$ and $\alpha ^{'}$  
of heavy
quarkonia is of great importance, since they afford opportunities  
for better
understanding of the dynamics of the strong interactions in the  
processes of
production of charmed and beauty hadrons at high energies.

Here we apply Regge phenomenology for the derivation of a mass  
formula for the
$SU(4)$ meson hexadecuplet, by assuming the (quasi)-linear
form of Regge trajectories for heavy quarkonia with slopes which are 
generally different from (less than) the standard one, $\alpha  
^{'}\simeq 0.9$
GeV$^{-2}.$ We show that for the formula to avoid depending on the  
values of
the slopes, it must be of sixth power in meson masses. It may be  
reduced to a
quadratic Gell-Mann--Okubo type relation, by fitting the values of  
the slopes,
which is in qualitative agreement with Eq. (5).

Let us assume the (quasi)-linear form of Regge trajectories for  
hadrons with
identical $J^{PC}$ quantum numbers (i.e., belonging to a common  
multiplet).
Then for the states with orbital momentum $\ell $ one has
\bqryn
\ell & = & \alpha ^{'}_{i\bar{i}}m^2_{i\bar{i}}\;+a_{i\bar{i}}(0), \\    
\ell & = & \alpha  
^{'}_{j\bar{i}}m^2_{j\bar{i}}\;\!+a_{j\bar{i}}(0), \\
\ell & = & \alpha ^{'}_{j\bar{j}}m^2_{j\bar{j}}+a_{j\bar{j}}(0).    
\eqryn
Using now the relation among the intercepts \cite{inter,Kai},
\beq
a_{i\bar{i}}(0)+a_{j\bar{j}}(0)=2a_{j\bar{i}}(0),
\eeq
one obtains from the above relations
\beq
\alpha ^{'}_{i\bar{i}}m^2_{i\bar{i}}+\alpha  
^{'}_{j\bar{j}}m^2_{j\bar{j}}=
2\alpha ^{'}_{j\bar{i}}m^2_{j\bar{i}}.
\eeq
In order to eliminate the Regge slopes from this formula, we need a  
relation
among the slopes. Two such relations have been proposed in the  
literature,
\beq
\alpha ^{'}_{i\bar{i}}\cdot \alpha ^{'}_{j\bar{j}}=\left( \alpha  
^{'}_{j\bar{i
}}\right) ^2,
\eeq
which follows from the factorization of residues of the $t$-channel poles
\cite{first,KY}, and
\beq
\frac{1}{\alpha ^{'}_{i\bar{i}}}+\frac{1}{\alpha  
^{'}_{j\bar{j}}}=\frac{2}{
\alpha ^{'}_{j\bar{i}}},
\eeq
based on topological expansion and the $q\bar{q}$-string picture of  
hadrons
\cite{Kai}.

For light quarkonia (and small differences in the $\alpha ^{'}$  
values), there
is no essential difference between these two relations; viz., for  
$\alpha ^{'
}_{j\bar{i}}=\alpha ^{'}_{i\bar{i}}/(1+x),$ $x\ll 1,$ Eq. (10)  
gives $\alpha
^{'}_{j\bar{j}}=\alpha ^{'}_{i\bar{i}}/(1+2x),$ whereas Eq. (9)  
gives $\alpha
^{'}_{j\bar{j}}=\alpha ^{'}_{i\bar{i}}/(1+x)^2\approx \alpha  
^{'}/(1+2x),$ i.e,
essentially the same result to order $x^2.$ However, for heavy  
quarkonia (and
expected large differences from the $\alpha ^{'}$ values for the light 
quarkonia) these relations are incompatible; e.g., for $\alpha  
^{'}_{j\bar{i}}=
\alpha ^{'}_{i\bar{i}}/2,$ Eq. (9) will give $\alpha  
^{'}_{j\bar{j}}=\alpha ^{
'}_{i\bar{i}}/4,$ whereas Eq. (10) $\alpha ^{'}_{j\bar{j}}=\alpha  
^{'}_{i\bar{
i}}/3.$ One has therefore to choose one of these relations in order  
to proceed
further. Here we use Eq. (10), since it is much more consistent  
with (8) than
is Eq. (9), which we tested by using measured quarkonia masses in  
Eq. (8). We
shall justify this choice in more detail in a separate publication  
\cite{prep}.

Since we are interested in $SU(4)$ breaking, and since it simplifies the 
discussion, we take average slope in the light quark sector:
\beq
\alpha ^{'}_{n\bar{n}}\cong \alpha ^{'}_{s\bar{n}}\cong \alpha  
^{'}_{s\bar{s}}
\cong \alpha ^{'}\simeq 0.85\;{\rm GeV}^{-2}.
\eeq
It then follows from the relations based on (8),
\bqry
\alpha ^{'}\eta ^2+\alpha ^{'}_{c\bar{c}}\eta _c^2 & = & 2\alpha  
^{'}_{c\bar{n
}}D^2, \\
\alpha ^{'}\eta _s^2+\alpha ^{'}_{c\bar{c}}\eta _c^2 & = & 2\alpha  
^{'}_{c
\bar{s}}D_s^2,
\eqry
that
\beq
\alpha ^{'}_{c\bar{n}}\cong \alpha ^{'}_{c\bar{s}}=\alpha  
^{'}\frac{(\eta _s^2-
\eta ^2)}{2(D_s^2-D^2)},
\eeq
\beq
\alpha ^{'}_{c\bar{c}}=\alpha ^{'}\left[ \frac{(\eta _s^2-\eta  
^2)}{(D_s^2-D^
2)}\frac{D^2}{\eta _c^2}-\frac{\eta ^2}{\eta _c^2}\right] .
\eeq
Using these values of the slopes in Eq. (10) with $i=n,\;j=c$, we obtain
\beq
\left( \eta _s^2D^2 - \eta ^2D_s^2\right) \left( \eta  
_s^2-\eta^2\right) +
\eta _c^2\left( D_s^2-D^2\right) \left( \eta _s^2-\eta ^2\right) =4\left(
\eta _s^2D^2-\eta ^2D_s^2\right) \left( D_s^2-D^2\right) ,
\eeq
which is a new mass relation for the $SU(4)$ meson hexadecuplet. To  
test (16),
we use the four well-established hexadecuplets\footnote{We use the value
$\eta _s=0.686$ GeV for pseudoscalar nonet, as follows from $\eta  
_s^2=2K^2-
\pi ^2.$ We also use the simple average of the masses of the  
isovector and
isoscalar states as the value of $\eta $ in Eqs. (16),(17) for the  
remaining
three multiplets.} \cite{data}:

1) 1 $^1S_0$ $J^{PC}=0^{-+},$ $m(\pi )=0.138$ GeV, $m(\eta  
_s)=0.686$ GeV,
$m(D)=1.868$ GeV, $m(D_s)=1.969$ GeV, $m(\eta _c)=2.980$ GeV,

2) 1 $^3S_1$ $J^{PC}=1^{--},$ $m(\rho )=0.775$ GeV, $m(\phi  
)=1.019$ GeV, $m(
D^\ast )=2.009$ GeV, $m(D^\ast _s)=2.112$ GeV, $m(J/\psi )=3.097$ GeV,

3) 1 $^1P_1$ $J^{PC}=1^{+-},$ $m(b_1)=1.201$ GeV,  
$m(h_1^{'})=1.380$ GeV, $m(D_
1)=2.422$ GeV, $m(D_{s1})=2.535$ GeV, $m(h_c(1P))=3.526$ GeV,

4) 1 $^3P_2$ $J^{PC}=2^{++},$ $m(a_2)=1.297$ GeV,  
$m(f_2^{'})=1.525$ GeV, $m(D_
2^\ast )=2.459$ GeV, $m(D_{s2}^\ast )=2.573$ GeV, $m(\chi  
_{c2}(1P))=3.556$
GeV,

and rewrite (16) in the form
\beq
\eta _c^2=\frac{(\eta _s^2D^2-\eta ^2D_s^2)\left(  
4(D_s^2-D^2)-(\eta _s^2-\eta
^2)\right) }{(\eta _s^2-\eta ^2)(D_s^2-D^2)},
\eeq
We shall test the relation by comparing the values of $\eta _c$  
given by Eq.
(17) with those established by experiment, using the known masses of the 
remaining states, for the four multiplets.

1) 1 $^1S_0$ $J^{PC}=0^{-+}.$ One obtains 3.137 GeV for the value  
of $m(\eta _
c)$ vs. experimentally established value 2.980 GeV; in this case  
the accuracy
of Eq. (16) is $\sim $ 5\%.

2) 1 $^3S_1$ $J^{PC}=1^{--}.$ In this case one obtains 3.202 GeV  
vs. 3.097 GeV,
as the value of $m(J/\psi );$ the accuracy is $\sim $ 3.5\%.

3) 1 $^1P_1$ $J^{PC}=1^{+-}.$ Now one obtains 3.615 GeV vs. 3.526  
GeV, as the
value of $m(h_c(1P));$ the accuracy is $\sim $ 2.5\%.

4) 1 $^3P_2$ $J^{PC}=2^{++}.$ In this case one has 3.618 GeV vs.  
3.556 GeV, as
the value of $m(\chi _{c2}(1P));$ the accuracy is $\sim $ 1.5\%.

One sees that the formula (16) holds with a high accuracy for all  
four well
established meson multiplets. The major contribution to the discrepancy 
between our formula result and experiment is the approximation  
(11). For higher
excited states the trajectories become more accurately parallel, and the 
approximation (11) and subsequent relation $\alpha  
^{'}_{c\bar{n}}=\alpha ^{'}
_{c\bar{s}} $ (Eq. (13)) become more exact. As shown above, the  
formula (16)
does hold with improving accuracy as one proceeds to higher spin  
multiplets.

A possible additional reason for the discrepancy is the curvature of the 
$\eta _c$ trajectory near $\ell =0$ since the mass of the $\eta _c$  
is lower
than expected from a linear extrapolation. Similarly, if one tries,  
apart from
its Goldstone nature, to fit the pion to the trajectory on which  
the $b_1(
1231)$ and $\pi _2(1670)$ lie $[\ell =0.85\;M^2(\ell )-0.30],$  
extrapolation
down to $\ell =0$ gives $m(\pi )\simeq 0.6$ GeV, much higher than  
the physical
value $m(\pi )=0.138$ GeV.

We note that a relation based on Eq. (9) for the slopes which is  
also of sixth
power in meson masses,
$$4D^2\left( D_s^2-D^2\right) \left( \eta _s^2-\eta ^2\right) =\eta _c^2
\left( \eta _s^2-\eta ^2\right) ^2+4\eta ^2\left( D_s^2-D^2\right) ^2,$$
holds for the four multiplets with an accuracy of 15-20\%. The  
reason for such
a large discrepancy with experiment is a lower value for the slope  
of the
charmonia trajectory given by Eq. (9), as compared to that given by  
(10),
leading to higher values for the charmonia masses.

We emphasize that the formula (16) does not depend on the $values$  
of the Regge
slopes, but only on the relation between them, Eq. (10), which  
justifies its
use in both the low energy region where the slopes are different  
and the high
energy region where all the slopes coincide. In the latter case, as  
follows
from (12),(13), $\eta _s^2-\eta ^2=2(D_s^2-D^2),$ and Eq. (17) reduces
to
\beq
\eta ^2+\eta _c^2=2D^2,
\eeq
consistent with Eq. (12) in this limit. One may also find from Eqs.  
(12),(13)
with equal slopes, and the standard $SU(3)$ Gell-Mann--Okubo relation,
\beq
\eta ^2+\eta _s^2=2K^2,
\eeq
that Eq. (3) also holds in this limit.

The entire analysis may, of course, be repeated with $B,B_s,\eta  
_b$ in place
of $D,D_s,\eta _c,$ respectively, and will lead (assuming (10)) to  
a relation
similar to (16):
\beq
\left( \eta _s^2B^2 - \eta ^2B_s^2\right) \left( \eta  
_s^2-\eta^2\right) +
\eta _b^2\left( B_s^2-B^2\right) \left( \eta _s^2-\eta ^2\right) =4\left(
\eta _s^2B^2-\eta ^2B_s^2\right) \left( B_s^2-B^2\right) ,
\eeq
which is a new mass relation for the $SU(5)$ meson 25-plet. We can  
test this
relation for vector mesons since the masses of all of the beauty states 
involved are established experimentally only for vector mesons  
\cite{data}:
$m(B^\ast )=5.325$ GeV, $m(B_s^\ast )=5.416$ GeV, $m(\Upsilon  
(1S))=9.460$ GeV.
The formula (20) yields $m(\Upsilon (1S))=9.796$ GeV, within $\sim  
$ 3.5\% of
the physical value, and the same accuracy as for charmed vector mesons. 

Let us now discuss the question of the generalization of the standard 
$SU(3)$ Gell-Mann--Okubo mass formula which is quadratic in mass to  
the case
of heavier quarkonia. We shall continue to assume the validity of  
Eq. (11)
and introduce $x>0$ through the relation
\beq
\alpha ^{'}_{c\bar{n}}=\alpha ^{'}_{c\bar{s}}=\frac{\alpha ^{'}}{1+x}.
\eeq
It then follows from (10) that
\beq
\alpha ^{'}_{c\bar{c}}=\frac{\alpha ^{'}}{1+2x},
\eeq
and one obtains from (12),(13),
\beq
(1+x)\left( \eta ^2+\eta _s^2\right) +\frac{2(1+x)}{1+2x}\eta  
_c^2=2\left( D^2
+D_s^2\right) .
\eeq
Results of the calculations of the Regge slopes of heavy quarkonia  
in refs.
\cite{KY}: $\alpha ^{'}_{c\bar{n}}/\alpha ^{'}\simeq \alpha  
^{'}_{c\bar{s}}/
\alpha ^{'}\simeq 0.73,$ $\alpha ^{'}_{c\bar{c}}/\alpha ^{'}\simeq  
0.58,$ and
\cite{Kai,CN}: $\alpha ^{'}_{c\bar{c}}\simeq 0.5$ GeV$^{-2},$  
support the value
\beq
x\cong 0.355.
\eeq
With this $x,$ it follows from (23) and the standard $SU(3)$  
Gell-Mann--Okubo
formula (19) that
\beq
8.13\;K^2+4.75\;\eta _c^2=6\left( D^2+D_s^2\right) .
\eeq

Thus, the new sixth order mass relations may be accurately reduced to
quadratic ones by use of specific values for the Regge slopes.

For the four well-established meson hexadecuplets, the formula (23)  
gives in
its l.h.s. and r.h.s., respectively\footnote{For the pseudovector  
nonet, we
use the value $K_{1B}=1.339$ GeV which follows from the assumption  
on a $45^o$
mixing between axial-vector and pseudovector nonets in the  
isodoublet channel
\cite{mix}.} (in GeV$^2):$

1) 1 $^1S_0$ $J^{PC}=0^{-+},\;\;\;$ 44.17 vs. 44.20

2) 1 $^3S_1$ $J^{PC}=1^{--},\;\;\;$ 52.03 vs. 50.98

3) 1 $^1P_1$ $J^{PC}=1^{+-},\;\;\;$ 73.63 vs. 73.75

4) 1 $^3P_2$ $J^{PC}=2^{++},\;\;\;$ 76.67 vs. 76.00

One sees that the formula (25) holds at a 1\% level for all four well 
established meson multiplets, thus confirming the assumption on the 
(quasi)-linearity of the Regge trajectories of heavy quarkonia in the low
energy region. The formula (25) is in qualitative agreement with  
the relation
(5) obtained by two of the present authors in ref. \cite{su4} by the 
application of the linear mass spectrum to a meson hexadecuplet.

Finally, we note that the derived Regge slopes in the charm sector are
\beq
\alpha ^{'}_{c\bar{n}}\simeq \alpha ^{'}_{c\bar{s}}\simeq  
0.63\;{\rm GeV}^{-2},
\;\;\;\alpha ^{'}_{c\bar{c}}\simeq 0.50\;{\rm GeV}^{-2}.
\eeq

\end{document}